\documentclass[letterpaper, 10 pt, conference]{ieeeconf}  

\IEEEoverridecommandlockouts                              
\usepackage{graphics} 
\usepackage{epsfig} 
\usepackage{mathptmx} 
\usepackage{times} 
\usepackage{amsmath} 
\usepackage{amssymb}  

\newtheorem{remark}{Remark}

\usepackage{algorithm}
\usepackage{algpseudocode}
\usepackage{stfloats}
\usepackage{graphicx}
\usepackage[caption=false,font=footnotesize]{subfig}
\usepackage{booktabs}

\title{\LARGE \bf
Local Sensitivity Analysis for Kernel-Regularized ARX Predictors in Data-Driven Predictive Control 
}

\author{Aihui Liu and Magnus Jansson
\thanks{This work was supported by VINNOVA Competence Center DigIT Lab Sustainable Industry, contract [2023-00556].}
\thanks{A. Liu and M. Jansson are with the Department of Information Science and Engineering, KTH Royal Institute of Technology, Stockholm, Sweden
{\tt\small \{aihui,janssonm\}@kth.se}. }%
}

\begin{document}

\maketitle
\thispagestyle{empty}
\pagestyle{empty}

\begin{abstract}
We study local sensitivity of structured ARX-based data-driven predictive control. Although predictor estimation is linear in the ARX parameters, the lifted multi-step predictor used in MPC depends on them implicitly, which complicates both uncertainty propagation and task-aware regularization. We derive a local first-order linearization of this implicit predictor map. The resulting Jacobian yields both an approximate control-relevant prediction uncertainty term and a task-dependent sensitivity metric for shaping kernel regularization. Numerical results show that the proposed analysis is most useful in weak-excitation regimes, where baseline SS regularization already provides substantial robustness gains and the proposed sensitivity shaping yields a further smaller improvement.
\end{abstract}

\section{Introduction}

Data-driven predictive control (DDPC) aims to reduce the reliance on an explicit state-space model by constructing the predictor directly from measured data. In recent years, DDPC has developed into a broad family of approaches with different predictor parameterizations \cite{handbook}. Regardless of the specific formulation, however, the closed-loop performance depends critically on the quality of the multi-step predictor embedded in the controller. In finite and noisy data regimes, predictor uncertainty therefore becomes a central determinant of closed-loop behavior.




From an identification-for-control perspective, the predictor should be assessed not only by how well it fits measured data, but also by how estimation uncertainty affects downstream multi-step prediction and control performance. Recent Bayesian, kernel-based, and bias-variance viewpoints \cite{BayesianKernel,coRe,bias_variance} all support this perspective. In the DDPC literature, this same concern has recently appeared in Final Control Error (FCE)-type formulations \cite{FCE}, where predictor uncertainty is incorporated directly into the control objective. Together, these viewpoints suggest that the important question is not only how to estimate predictor parameters accurately, but also how to characterize and shape estimation error in directions that matter for the control objective.




In structured ARX-based DDPC, a key difficulty remains: while the identification step is linear in the predictor parameters, the lifted multi-step predictor used inside MPC depends on them implicitly and nonlinearly. This mismatch is the main obstacle to propagating parameter uncertainty into a control-relevant multi-step prediction analysis.


To address this, we derive a local first-order sensitivity analysis of the implicit ARX multi-step predictor around a nominal estimate. The resulting Jacobian provides a tractable map from predictor-parameter perturbations to control-weighted multi-step prediction perturbations. It is then used in two ways: first, to approximate an FCE-type uncertainty term in the MPC objective; second, to construct a task-dependent regularization metric for predictor estimation. The focus of this paper is to build a practical local bridge between parameter uncertainty and control-relevant predictor sensitivity, and to demonstrate where this bridge is useful.

The main contributions of this paper are:




(i) We derive a local first-order sensitivity analysis for the implicit multi-step predictor in structured ARX-based DDPC, bridging linear predictor identification and control-relevant multi-step uncertainty propagation.

(ii) We show that the same Jacobian has two uses: it yields an approximate FCE-type uncertainty term and a task-dependent sensitivity metric for shaping kernel regularization.

(iii) We show numerically that the main practical value appears under weak excitation: baseline SS regularization substantially improves robustness, and the proposed sensitivity shaping gives an additional improvement over the baseline SS prior.


This paper is organized as follows. Section \ref{sec:ARX} introduces the structured ARX-based multi-step predictor and its predictor-parameter representation. Section \ref{sec:linearize} develops a local linearization of the implicit predictor map and uses it to propagate posterior parameter uncertainty into an approximate FCE-type cost. Section \ref{sec:ker_co} presents kernel-regularized identification and shows how the same local sensitivity induces a task-aware sensitivity-based kernel. Section \ref{sec:experiment} reports numerical simulations, and Section \ref{sec:conclusion} concludes the paper and discusses future work.

\section{Structured ARX Multi-Step Predictor} \label{sec:ARX}

\subsection{Predictor Markov parameters and ARX representation}

We consider the discrete-time linear time-invariant (LTI) system in innovations form:
\begin{equation} \label{eq:inno_form}
\begin{cases}
x(t+1) = A x(t) + B u(t) + K e(t), \\
y(t) = C x(t) + D u(t) + e(t), 
\end{cases}
\end{equation}
where $x(t) \in \mathbb{R}^{n},u(t) \in \mathbb{R}^{n_u}$ and $y(t) \in \mathbb{R}^{n_y}$ denote the state, input and output, respectively, and $e(t)$ is a zero-mean white innovation process.  

It is convenient to rewrite (\ref{eq:inno_form}) by introducing $\tilde{A}=A-KC$ and $\tilde{B}=B-KD$, which gives
\begin{equation} \label{eq:pred_form}
\begin{cases}
x(t+1) = \tilde{A} x(t) + \tilde{B} u(t) + K y(t), \\
y(t) = C x(t) + Du(t)+ e(t) .
\end{cases} 
\end{equation}

For a chosen past horizon $L_p$ and future horizon $L_f$, define the stacked past and future signals 
\begin{align}
    \mathbf{u}_p(t) &:= \text{col}(u(t-L_p), u(t-L_p+1), \dots, u(t-1) ),  \\
    \mathbf{y}_p(t) &:= \text{col}(y(t-L_p), y(t-L_p+1), \dots, y(t-1) ),  \\
    \mathbf{u}_f(t) &:= \text{col}(u(t), u(t+1), \dots, u(t+L_f-1) ), \\
    \mathbf{y}_f(t) &:= \text{col}(y(t), y(t+1), \dots, y(t+L_f-1) ) ,
\end{align}
where $\text{col}(\cdot)$ denotes the column vector. Similarly, 
\begin{equation}
    \mathbf{e}_f(t) := \text{col}( e(t), e(t+1), \dots, e(t+L_f-1) ).
\end{equation}
We also use the notation for the past data $\mathbf{z}_p(t):=[\mathbf{y}^\top_p(t), \mathbf{u}^\top_p(t)]^\top$.

The predictor form in (\ref{eq:pred_form}) induces a one-step-ahead input-output predictor that can be written as
\begin{equation} \label{eq:ARX_inf}
y(t) \approx \sum_{i=1}^{\infty} \phi_{y}^{i} y(t-i) + \sum_{j=0}^{\infty} \phi_{u}^{j}u(t-j) + e(t),
\end{equation}
where the matrices $\phi_{u}^0=D$, $\phi_{u}^j=C\tilde{A}^{j-1} \tilde{B}$ for $j \geq 1$, and $\phi_{y}^i=C\tilde{A}^{i-1} K$ for $i \geq 1$ are called the ARX coefficients or the predictor Markov parameters. 

In practice, we truncate this predictor and estimate a finite-order ARX model
\begin{equation} \label{eq:arx}
y(t) \approx \sum_{i=1}^{n_a} \hat{\phi}_{y}^{i} y(t-i) + \sum_{j=0}^{n_b} \hat{\phi}_{u}^{j}u(t-j) + e(t),
\end{equation}

For identification, we collect all predictor parameters in 
\begin{equation} \label{eq: theta}
    \theta := \text{col} (\phi_y^1, \dots,\phi_y^{n_a}, \phi_u^0, \dots,\phi_u^{n_b}) ,
\end{equation}
define $\mathbf{y}:=\text{col}(y(t), \dots,y(t+N-1))$, $\mathbf{e}:=\text{col}(e(t), \dots,e(t+N-1))$, and arrange (\ref{eq:arx}) row-wise in vector form as 
\begin{equation} \label{eq:arx_regress}
    \mathbf{y} = H\theta +\mathbf{e}
\end{equation}
where $H$ is the regressor Hankel assembled from the corresponding past signals.

The key advantage of this parameterization is that identification is linear in $\theta$, so posterior means and covariances remain available in closed form under regularization. Also, under the Gaussian noise assumption, the data enter the estimator through the sufficient statistics $H^\top H$ and $H^\top \mathbf{y}$. 


\subsection{ARX-Based Multi-Step Predictor}

We can write the state equation in (\ref{eq:pred_form}) sequentially as 
\begin{equation} \label{eq:est_x}
    x(t) = \sum_{k=0}^{L_p-1} \tilde{A}^k [ K y(t-k-1) + \tilde{B} u(t-k-1) ]  + \tilde{A}^{L_p} x(t-L_p)
\end{equation}
For sufficiently large $L_p$ and stable $\tilde{A}$, the term $\tilde{A}^{L_p}$ is neglected, yielding an approximate dependence of the predictor state on past data. The resulting multi-step predictor can be written as
\begin{equation} \label{eq:multi_predictor}
    \mathbf{y}_f = \begin{bmatrix}
        \Psi_u & \Psi_y
    \end{bmatrix} \begin{bmatrix}
        \mathbf{u}_p \\ \mathbf{y}_p
    \end{bmatrix} + \Phi_u \mathbf{u}_f + \Phi_y \mathbf{y}_f+\mathbf{e}_f 
\end{equation}
The matrices $\Psi_u, \Psi_y, \Phi_u,\Phi_y$ in this multi-step predictor are block Toeplitz matrices assembled from the ARX coefficient sequences $\{\phi_u^j\}$ and $\{\phi_y^i\}$:
\begin{equation} \label{eq:four}
\begin{aligned}
    \Psi_u &= \begin{bmatrix}
        \phi^{L_p}_u & \dots & \phi_u^2 & \phi_u^1 \\
        \phi^{L_p+1}_u & \dots & \phi_u^3 & \phi_u^2 \\
        \vdots & & \vdots & \vdots \\
        \phi^{L_p+L_f-1}_u & \dots & \phi^{L_f+1}_u & \phi^{L_f}_u
    \end{bmatrix} , \\
    \Psi_y &= \begin{bmatrix}
        \phi^{L_p}_y & \dots & \phi_y^2 & \phi_y^1 \\
        \phi^{L_p+1}_y & \dots & \phi_y^3 & \phi_y^2 \\
        \vdots & & \vdots & \vdots \\
        \phi^{L_p+L_f-1}_y & \dots & \phi^{L_f+1}_y & \phi^{L_f}_y
    \end{bmatrix}, \\
    \Phi_u &= \begin{bmatrix}
        \phi_u^0 & & & \\
        \phi_u^1 & \phi_u^0 & & \\
        \vdots & \vdots & \ddots & \\
        \phi^{L_f-1}_u & \phi^{L_f-2}_u & \dots & \phi_u^0
    \end{bmatrix} , \\
    \Phi_y &= \begin{bmatrix}
        0 & & & \\
        \phi_y^1 & 0 & & \\
        \vdots & \vdots & \ddots & \\
        \phi^{L_f-1}_y & \phi^{L_f-2}_y & \dots & 0 .
    \end{bmatrix}
\end{aligned}
\end{equation}

Typically, the ARX orders $n_a, n_b \leq L_p$ are chosen no larger than $L_p$. Practically we can choose the ARX order by AIC or BIC, and then construct the matrices in (\ref{eq:four}) by padding the higher-order ARX coefficients with zeros as needed. Note that in (\ref{eq:est_x}), we are ignoring higher-order terms of $\tilde{A}^{L_p}$. Accordingly, in (\ref{eq:four}), coefficients beyond the retained past horizon are zero-padded, i.e., terms starting from $\phi^{L_p}$ onward are neglected in this approximation.

The linear regression equation (\ref{eq:multi_predictor}) expresses the future trajectory through past data, future inputs, and future outputs. This structured parameterization is more compact than a trajectory-coefficient description and is directly compatible with kernel priors and posterior covariance calculations.
\begin{equation} \label{eq:predictor}
    \hat{\mathbf{y}}_f(\mathbf{u}_f;\theta) = (I-\Phi_y)^{-1} \left( \begin{bmatrix}
        \Psi_u & \Psi_y
    \end{bmatrix} \begin{bmatrix}
        \mathbf{u}_p \\ \mathbf{y}_p
    \end{bmatrix} + \Phi_u \mathbf{u}_f \right)
\end{equation}

Equation (\ref{eq:predictor}) is the central object used by the controller. It is affine in the lifted signals but nonlinear in the identified parameter vector, since $\theta$ enters both the affine term and the inverse $(I-\Phi_y)^{-1}$. This is the source of the uncertainty-propagation difficulty addressed in the next section.





\section{Local Linearization and Uncertainty Propagation} \label{sec:linearize}

This section develops the main technical bridge of the paper: a local map from predictor-parameter uncertainty to control-relevant multi-step prediction sensitivity.

Let $(\bar{\theta}, \Sigma_\theta)$ denote the nominal estimate of the predictor parameters for (\ref{eq: theta}), and its corresponding posterior covariance. In this section, we focus on how this parameter uncertainty $\Sigma_\theta$ propagates into control-relevant output uncertainty.

\subsection{Local linearization of the implicit predictor}

The multi-step predictor (\ref{eq:predictor}) can be written as 
\begin{equation} \label{eq:pre_ab}
    \hat{\mathbf{y}}_f(\mathbf{u}_f;\theta) = A(\theta)^{-1}  \Big( b(\theta) +  \Phi_u(\theta) \mathbf{u}_f \Big)
\end{equation}
where 
\begin{equation}
    A(\theta) :=  I-\Phi_y(\theta) , \quad b(\theta) :=  \Psi_u(\theta) \mathbf{u}_p + \Psi_y(\theta) \mathbf{y}_p
\end{equation}
To propagate parameter uncertainty analytically, we locally linearize the predictor around the estimated parameter $\bar{\theta}$. 

For a fixed candidate control sequence $u_f$, we linearize the predictor map locally around $\bar{\theta}$
\begin{equation} \label{eq:local}
    \hat{\mathbf{y}}_f(\mathbf{u}_f;\theta) \approx \hat{\mathbf{y}}_f(\mathbf{u}_f;\bar{\theta}) + J(\bar{\theta},\mathbf{u}_f) (\theta-\bar{\theta}) 
\end{equation}
where 
\begin{equation} \label{eq:jacobian}
    J (\bar{\theta},\mathbf{u}_f) := \frac{\partial \hat{\mathbf{y}}_f(\mathbf{u}_f;\theta)}{\partial \theta} \Big \vert _{\theta = \bar{\theta}} \in \mathbb{R}^{n_yL_f \times n_{\theta}} .
\end{equation}

To compute this Jacobian, the main technical step is to differentiate through the implicit inverse. Note that
\begin{equation}
\begin{aligned}
    dA(\theta) & = -d\Phi_y(\theta), \\
    d(A(\theta)^{-1})  & = -A(\theta)^{-1}(dA(\theta))A(\theta)^{-1} ,
\end{aligned}
\end{equation}
hence,
\begin{equation}
\begin{aligned}
    d\hat{\mathbf{y}}_f & = d \Big(A^{-1}(b+\Phi_u \mathbf{u}_f) \Big) \\
    & = (dA^{-1})(b+\Phi_u \mathbf{u}_f) + A^{-1}(db+d\Phi_u \mathbf{u}_f)  \\
    & = A^{-1}\Big(db+d\Phi_u \mathbf{u}_f + (d\Phi_y) \hat{\mathbf{y}}_f\Big) 
\end{aligned}
\end{equation}
The derivative separates naturally into three effects: perturbations entering a past-data term, a future-input term, and a future-output feedback term.

For differentiation, we use $\theta$ to denote the padded coefficient vector, with a slight abuse of notation.
\begin{equation}
    \theta = \text{col} (\phi_y^1, \dots,\phi_y^{L_p+L_f-1}, \phi_u^0, \dots,\phi_u^{L_p+L_f-1})
\end{equation}
which can fully parametrize the matrices. For differentiation, we temporarily use the padded coefficient vector that parameterizes all entries appearing in the lifted matrices; lower-order models are recovered by zero padding. 

Since $\Psi_u, \Psi_y, \Phi_u, \Phi_y$ depend linearly on the common coefficient vector $\theta$, their derivatives can be represented by fixed coefficient-placement matrices
\begin{equation}
    \left\{  E^{\Psi_u}_i, E^{\Psi_y}_i, E^{\Phi_u}_i, E^{\Phi_y}_i  \right\}^{n_{\theta}}_{i=1}
\end{equation}
such that 
\begin{equation} \label{eq:Ematrix}
\begin{aligned}
    \Psi_u(\theta) = \sum_{i=1}^{n_\theta} \theta_{i} E^{\Psi_u}_i, \quad 
    \Psi_y(\theta) = \sum_{i=1}^{n_\theta} \theta_{i} E^{\Psi_y}_i,  \\
    \Phi_u(\theta) = \sum_{i=1}^{n_\theta} \theta_{i} E^{\Phi_u}_i, \quad 
    \Phi_y(\theta) = \sum_{i=1}^{n_\theta} \theta_{i} E^{\Phi_y}_i. \\
\end{aligned}
\end{equation}

From (\ref{eq:pre_ab}) and (\ref{eq:Ematrix}), 
\begin{equation}
    \frac{\partial b}{\partial \theta_i} = E^{\Psi_u}_i \mathbf{u}_p + E^{\Psi_y}_i \mathbf{y}_p, \; \frac{\partial \Phi_u}{\partial \theta_i} =E^{\Phi_u}_i, \; \frac{\partial \Phi_y}{\partial \theta_i} =E^{\Phi_y}_i .
\end{equation}
Because all lifted matrices depend linearly on the same coefficient vector, the Jacobian can be expressed through fixed coefficient-placement matrices. Therefore, take the $i$-th column of the Jacobian and evaluate at $\theta=\bar{\theta}$:
\begin{equation}\label{eq:Ji}
\begin{aligned}
J_i(\bar{\theta},\mathbf{u}_f)
&= A(\bar{\theta})^{-1} \Big(
E^{\Psi_u}_i \mathbf{u}_p
+ E^{\Psi_y}_i \mathbf{y}_p
+ E^{\Phi_u}_i \mathbf{u}_f  \\
&\qquad\qquad
+ E^{\Phi_y}_i \hat{\mathbf{y}}_f(\mathbf{u}_f;\bar{\theta})
\Big)
\end{aligned}
\end{equation}
and thus 
\begin{equation}
    J(\bar{\theta}, \mathbf{u}_f) = [J_1(\bar{\theta}, \mathbf{u}_f) \; \dots \; J_{n_\theta}(\bar{\theta}, \mathbf{u}_f)] .
\end{equation}

Equation (\ref{eq:Ji}) shows a perturbation in $\theta_i$ affects the predicted output through three contributions: the past data term $\Psi_u, \Psi_y$, the explicit future-input term $\Phi_u \mathbf{u}_f$, and the implicit future-output feedback term $\Phi_y \hat{\mathbf{y}}_f$. This decomposition will be used next to separate the dependence of the Jacobian on the decision variable $\mathbf{u}_f$.


\subsection{Affine dependence on the future input}

For later use in the Final Control Error (FCE)-type cost, it is convenient to make the dependence of the Jacobian on $\mathbf{u}_f$ explicit. At the expansion point $\bar{\theta}$, define
\begin{equation}
    \bar{A}:=A(\bar{\theta}), \quad \bar{b}:=b(\bar{\theta}), \quad \bar{\Phi}_u:= \Phi_u(\bar{\theta}),
\end{equation}
so that 
\begin{equation} \label{eq:ythetabar}
    \hat{\mathbf{y}}_f(\mathbf{u}_f;\bar{\theta}) = \bar{A}^{-1}\bar{b} +  \bar{A}^{-1} \bar{\Phi}_u \mathbf{u}_f
\end{equation}

Substituting the nominal predictor expression (\ref{eq:ythetabar}) into (\ref{eq:Ji}) makes the dependence of each Jacobian column on $\mathbf{u}_f$ explicit:
\begin{equation} 
\begin{aligned}
    J_i(\bar{\theta}, \mathbf{u}_f) =&  \bar{A}^{-1} \left( E^{\Psi_u}_i \mathbf{u}_p +  E^{\Psi_y}_i \mathbf{y}_p +E^{\Phi_y}_i \bar{A}^{-1}\bar{b}  \right) \\
    & + \bar{A}^{-1} \left(  E^{\Phi_u}_i + E^{\Phi_y}_i  \bar{A}^{-1} \bar{\Phi}_u  \right) \mathbf{u}_f
\end{aligned}
\end{equation}

In particular, each column can be separated into a term that depends only on the current operating condition and a term that depends affinely on the candidate future input $\mathbf{u}_f$. Define the $i$-th column of $J$ as
\begin{equation}
    J_i(\bar{\theta}, \mathbf{u}_f) = J^0_i + J^1_i \mathbf{u}_f, \quad i=1,\dots, n_\theta, 
\end{equation}
where $J_i^0 \in \mathbb{R}^{n_y L_f}$ and $J_i^1 \in \mathbb{R}^{n_y L_f \times n_u L_f}$ are:
\begin{equation}
\begin{aligned}
    J^0_i & := \bar{A}^{-1} \left( E^{\Psi_u}_i \mathbf{u}_p +  E^{\Psi_y}_i \mathbf{y}_p +E^{\Phi_y}_i \bar{A}^{-1}\bar{b}  \right) \\
    J^1_i & := \bar{A}^{-1} \left(  E^{\Phi_u}_i + E^{\Phi_y}_i  \bar{A}^{-1} \bar{\Phi}_u  \right) .
\end{aligned}
\end{equation}
Hence, each Jacobian column depends affinely on $\mathbf{u}_f$, and the full Jacobian $J(\bar{\theta}, \mathbf{u}_f)$ is columnwise affine in $\mathbf{u}_f$.

This affine dependence is important because it allows the uncertainty contribution to inherit an explicit dependence on the decision variable $\mathbf{u}_f$. Since then it can be incorporated into the MPC objective, as in \cite{FCE}.

\subsection{Uncertainty propagation}

A first consequence of the local linearization is an approximate output covariance and an associated FCE-type uncertainty penalty. Assume now that, after identification from data $\mathcal{D}$, the parameter posterior is approximated by
\begin{equation}
    \theta | \mathcal{D} \sim \mathcal{N} (\bar{\theta}, \Sigma_\theta)
\end{equation}
where $\bar{\theta}$ is typically the posterior mean. Under the local approximation (\ref{eq:local}),
\begin{equation}
    \hat{\mathbf{y}}_f(\mathbf{u}_f; \theta) -\hat{\mathbf{y}}_f(\mathbf{u}_f; \bar{\theta}) \approx J(\bar{\theta}, \mathbf{u}_f) (\theta-\bar{\theta}) .
\end{equation}
This local approximation yields an explicit approximation of the predicted output covariance and, therefore, of the additional uncertainty term in the control cost. 

The conditional covariance of $\mathbf{y}_f$ is approximated by
\begin{equation}
    \Sigma_{\mathbf{y}_f}(\mathbf{u}_f) := \text{cov} [\hat{\mathbf{y}}_f(\mathbf{u}_f; \theta)|\mathcal{D}] \approx J(\bar{\theta}, \mathbf{u}_f) \Sigma_\theta J(\bar{\theta}, \mathbf{u}_f)^\top
\end{equation}

For the reference signal $\mathbf{r}_f \in \mathbb{R}^{n_y L_f}$ and the quadratic weight $Q$ and $R$, the nominal MPC tracking cost is
\begin{equation}
    L_{\text{MPC}}(\mathbf{u}_f) = \Vert \mathbf{r}_f - \hat{\mathbf{y}}_f(\mathbf{u}_f; \theta) \Vert_{Q}^2 + \Vert  \mathbf{u}_f \Vert_{R}^2 
\end{equation}

One immediate use of the local linearization is to approximate how posterior parameter uncertainty contributes an additional term to the MPC objective. Following the FCE formulation in \cite{FCE}, we define
\begin{equation} \label{eq:FCE}
\begin{aligned}
    & L_{\text{FCE}}(\mathbf{u}_f)  = \mathbb{E}_\theta \left[ L_{\text{MPC}}(\mathbf{u}_f) | \mathcal{D} \right] \\
     =&  \mathbb{E} \left[ \Vert \mathbf{r}_f -\hat{\mathbf{y}}_f(\mathbf{u}_f;\bar{\theta})+ \hat{\mathbf{y}}_f(\mathbf{u}_f;\bar{\theta}) - \hat{\mathbf{y}}_f(\mathbf{u}_f;\theta)  \Vert_{Q}^2 + \Vert  \mathbf{u}_f \Vert_{R}^2 | \mathcal{D} \right] \\
    =& \Vert \mathbf{r}_f -\hat{\mathbf{y}}_f(\mathbf{u}_f;\bar{\theta}) \Vert_Q^2  + \Vert  \mathbf{u}_f \Vert_{R}^2 + \mathbb{E} \left[  \Vert \hat{\mathbf{y}}_f(\mathbf{u}_f;\theta) - \hat{\mathbf{y}}_f(\mathbf{u}_f;\bar{\theta}) \Vert^2_Q |\mathcal{D} \right] \\
    & \quad + 2(\mathbf{r}_f -\hat{\mathbf{y}}_f(\mathbf{u}_f;\bar{\theta}))^\top Q\mathbb{E} \left[   (\hat{\mathbf{y}}_f(\mathbf{u}_f;\theta) - \hat{\mathbf{y}}_f(\mathbf{u}_f;\bar{\theta}))  | \mathcal{D} \right]
\end{aligned}
\end{equation}
Following \cite{FCE}, we neglect the cross term within the local approximation.

Under this approximation, the difference between the nominal MPC cost $L_{\text{MPC}}$ and the FCE-type cost $L_{\text{FCE}}$ is the uncertainty term $L_\Omega$. The local approximation gives it in closed form:
\begin{equation} \label{eq:Omega}
\begin{aligned}
    & L_\Omega(\mathbf{u}_f) = \mathbb{E} \left[  \Vert \hat{\mathbf{y}}_f(\mathbf{u}_f;\theta) - \hat{\mathbf{y}}_f(\mathbf{u}_f;\bar{\theta}) \Vert^2_Q |\mathcal{D} \right]  \\
    & \approx \mathbb{E} \left[  (\theta-\bar{\theta})^\top J^\top Q J (\theta-\bar{\theta}) \right] = \operatorname{tr} \left( J^\top   Q J \Sigma_\theta \right)
\end{aligned}
\end{equation}
Using the linearized predictor, this term becomes a quadratic form in $\theta-\bar{\theta}$, whose expectation is the trace expression in (\ref{eq:Omega}). Hence, the local Jacobian converts posterior parameter covariance into an explicit control-relevant penalty. The resulting approximate objective is given by
\begin{equation} 
    L_{\text{FCE}}(\mathbf{u}_f) \approx L_{\text{MPC}}(\mathbf{u}_f) + L_\Omega(\mathbf{u}_f)
\end{equation}

\begin{remark}
    The vanishing of the cross term in (\ref{eq:FCE}) in \cite{FCE} is only justified within the local approximation and should therefore be interpreted as approximate, especially when kernel regularization introduces bias.
\end{remark}

In the present paper, this uncertainty term is mainly used to motivate the sensitivity metric developed next; its empirical impact as a direct MPC cost augmentation will be assessed separately in Section \ref{sec:experiment}.

In summary, the local linearization yields an explicit first-order map from posterior parameter covariance to weighted multi-step prediction uncertainty. In the next section, we use the same Jacobian to construct a task-dependent regularization term.


\section{Kernel-Regularized Identification and Sensitivity-based Kernel Design} \label{sec:ker_co}

In this Section, we use kernel-regularized ARX identification to estimate the nominal $\bar{\theta}$ and posterior covariance $\Sigma_\theta$. We also specify how the local sensitivity derived in Section \ref{sec:linearize} is used to shape a task-aware kernel.

\subsection{Kernel-Regularized Identification} \label{sec:kernel}

Although the multi-step predictor used by MPC is nonlinear in $\theta$, the ARX identification step remains linear, so Gaussian priors and posterior uncertainty are available in closed form. The ARX linear regression in (\ref{eq:arx_regress}) is:
\begin{equation} 
    \mathbf{y} = H\theta +\mathbf{e} ,
\end{equation}
where the regressor $H$ is built by stacking past input and output data. 

Following the classic kernel identification literature \cite{Pillonetto2010,kernelsurvey}, we place a Gaussian prior on $\theta$,
\begin{equation}
    \theta \sim \mathcal{N}(0,K(\eta)),
\end{equation}
where $K(\eta) \succ 0$ is a kernel covariance matrix with hyperparameters $\eta$. In practice, $K(\eta)$ may be chosen from standard TC/SS-type kernel families, and can be block-structured to distinguish the output-side and input-side predictor coefficients, e.g. 
\begin{equation} \label{eq:K}
    K(\eta) = \begin{bmatrix}
        K_y(\eta_y) & 0 \\
        0 & K_u(\eta_u)
    \end{bmatrix} .
\end{equation}
This yields a structured prior directly on the predictor Markov parameters.

Assume the regression noise to be white with covariance $\sigma^2 I_{n_\theta}$, the kernel-regularized estimator is 
\begin{equation} \label{eq:theta_ss}
    \hat{\theta}(\eta) = \arg \min_\theta  \frac{1}{\sigma^2} \Vert \mathbf{y}-H\theta \Vert^2 + \Vert \theta \Vert^2_{K(\eta)^{-1}},
\end{equation}
with closed-form solution
\begin{equation}
    \hat{\theta}(\eta) = \left( \frac{1}{\sigma^2} H^\top H +  K(\eta)^{-1} \right)^{-1} \frac{1}{\sigma^2} H^\top  \mathbf{y} .
\end{equation}

Under the Gaussian model
\begin{equation}
    \mathbf{y}|\theta \sim \mathcal{N}(H\theta, \sigma^2 I),
\end{equation}
the posterior is also Gaussian:
\begin{equation}
    \theta |\mathbf{y}, \eta \sim \mathcal{N}(\bar{\theta}, \Sigma_\theta),
\end{equation}
where 
\begin{equation}
    \Sigma_\theta (\eta) = \left( \frac{1}{\sigma^2} H^\top H +  K(\eta)^{-1} \right)^{-1}, \; \bar{\theta}(\eta) = \frac{1}{\sigma^2}\Sigma_\theta (\eta) H^\top  \mathbf{y} .
\end{equation}

The TC/SS kernel hyperparameters are selected by Empirical Bayes (EB). Marginalizing out $\theta$ gives
\begin{equation}
    \mathbf{y} \sim\mathcal{N}(0, \Sigma_{\mathbf{y}}(\eta,\sigma^2)), \;  \Sigma_{\mathbf{y}}(\eta, \sigma^2) = HK(\eta)H^\top+\sigma^2 I,
\end{equation}
and therefore
\begin{equation}
    \hat{\eta}, \hat{\sigma}^2 = \arg \min_{\eta, \sigma^2} \log \det \Sigma_\mathbf{y}(\eta, \sigma^2) +  \mathbf{y}^\top \Sigma_\mathbf{y}(\eta, \sigma^2)^{-1} \mathbf{y} . 
\end{equation}

All estimation steps are implemented using Cholesky factorizations rather than explicit matrix inversion. This allows stable evaluation of posterior means, quadratic forms, and log-determinants through triangular solves and diagonal entries.


This baseline kernel-regularized estimator provides both the nominal predictor and the posterior covariance needed for the local sensitivity analysis.




\subsection{Sensitivity-based Kernel Design} \label{sec:controlOri}

The main practical use of the local Jacobian in this paper is to construct a task-dependent sensitivity metric that complements the baseline identification-oriented prior.


From the uncertainty propagation analysis in Section \ref{sec:linearize}, the additional weighted prediction uncertainty is approximated by (\ref{eq:Omega}):
\begin{equation}
\begin{aligned}
    L_\Omega(\mathbf{u}_f)  & = \mathbb{E} \left[  \Vert \hat{\mathbf{y}}_f(\theta) - \hat{\mathbf{y}}_f(\bar{\theta}) \Vert^2_Q |\mathcal{D}, \xi \right]  \\
     & \approx  \operatorname{tr} \left( J(\bar{\theta},\xi)^\top   Q J(\bar{\theta},\xi) \Sigma_\theta \right) 
\end{aligned}
\end{equation}
where $\xi$ denotes the current task or operating condition, including the relevant past data $\mathbf{u}_p, \mathbf{y}_p$ and the nominal future input $\mathbf{u}_f$ used to evaluate local prediction sensitivity for the target task. $J(\bar{\theta}, \xi)$ is the Jacobian of the multi-step predictor with respect to $\theta$ as in (\ref{eq:jacobian}):
\begin{equation}
    J (\bar{\theta},\xi) := \frac{\partial \hat{\mathbf{y}}_f(\theta, \xi)}{\partial \theta} \Big \vert _{\theta = \bar{\theta}} . 
\end{equation}
$J (\bar{\theta},\xi)$ is a local mapping that depends on the $\xi$, i.e., the current state or the past data and potential future control input.

The uncertainty analysis suggests the local sensitivity matrix
\begin{equation} \label{eq:W}
    W(\xi):=  J (\bar{\theta},\xi)^\top Q  J (\bar{\theta},\xi) .
\end{equation}
which depends on the current operating condition $\xi$. $\xi$ includes the relevant past data $\mathbf{u}_p, \mathbf{u}_f$ and the nominal future input $\mathbf{u}_f$ associated with the target task. For a local perturbation $\delta \theta$, the quantity $\delta\theta^\top W(\xi) \delta \theta$ measures the first-order weighted prediction effect of that perturbation. The large value of $\delta\theta^\top W(\xi) \delta \theta$ produces a large increase in the weighted control-relevant prediction error, and those directions with small value are comparatively benign. This motivates using $W(\xi)$ to shape the regularization more strongly along directions that are locally more influential for the downstream control objective.





Since the local sensitivity depends on the operating condition, we average it over representative task realizations $\xi_{\text{task}}$ near the target regime. The averaged sensitivity matrix yields:
\begin{equation}
    \bar{W}:= \mathbb{E}_{\xi_{\text{task}}} \left[  J (\bar{\theta},\xi_{\text{task}})^\top Q  J (\bar{\theta},\xi_{\text{task}})  \right] ,
\end{equation}
and in practice, it can be approximated by 
\begin{equation} \label{eq:Wbar}
    \bar{W} \approx \frac{1}{N_{\text{task}}} \sum_{j=1}^{N_{\text{task}}} J (\bar{\theta},\xi_{j})^\top Q  J (\bar{\theta},\xi_{j})
\end{equation}
where $\{ \xi_j \}_{j=1}^{N_\text{task}}$ at each time step $j$ are extracted from the task-specific operating data. Thus, $\bar{W}$ captures which parameter directions matter on average for the training tasks of interest. We can normalize it for better numerics:
\begin{equation} \label{eq:normW}
    \bar{W}_{\text{norm}} = \frac{\bar{W}}{\frac{1}{n_\theta} \operatorname{tr}(\bar{W})}.
\end{equation}
and use it to shape regularization beyond the baseline TC/SS prior. 


The proposed kernel shaping is intended as a local refinement of the baseline identification-oriented kernel from Section \ref{sec:kernel}. Let $K$ denote the fixed baseline kernel covariance obtained after the Empirical Bayes tuning in the ARX identification step. Using this kernel, the second-stage estimator is
\begin{equation} 
\hat\theta_c = \arg\min_{\theta} \frac{1}{\sigma^2}\| \mathbf{y}-H\theta\|_2^2 + \theta^\top K^{-1}\theta + \mu(\theta-\bar{\theta})^\top \bar W (\theta-\bar{\theta}) .
\label{eq:co_estimator}
\end{equation}
The shaped kernel therefore penalizes parameter variation more strongly in directions that are locally important for the downstream control objective, while preserving the computational simplicity of quadratic regularization.



The resulting estimator remains available in closed form:
\begin{equation} \label{eq:co_closed_form}
\hat\theta_c = \left(
\frac{1}{\sigma^2}H^\top H+ K^{-1}+\mu \bar W \right)^{-1} \left( \frac{1}{\sigma^2}H^\top \mathbf{y} + \mu \bar{W}\bar{\theta} \right).
\end{equation}
This can be interpreted as combining the baseline prior precision with an additional sensitivity-based precision term.


\begin{remark} \label{rmk:twostage}
    In practice, when task-relevant data is not available, one convenient realization is a two-stage workflow. First, we use TC/SS kernel to identify the nominal $\bar{\theta}$, build the controller, and collect representative operating data for the target task. Then we compute the control-oriented kernel, re-identify, and build the new controller. When task-relevant data are already available a priori, $\bar{W}$ can be constructed directly.
\end{remark}


\begin{remark}
In this paper, after the baseline ARX kernel-identification step, we fix $\sigma^2$ and the baseline kernel $K$ obtained from EB tuning, and tune the sensitivity kernel hyperparameter $\mu$ manually. Other possibilities in hyper-parameter tuning are left for future work.
\end{remark}


\begin{remark}
The same Jacobian-based construction could in principle be used with other differentiable predictor parameterizations, such as Fundamental Lemma-based predictors. But the number of parameters may become prohibitive.
\end{remark}

\section{Experiments} \label{sec:experiment}

We evaluate the proposed framework in two identification regimes with different levels of informativeness and excitation. The first is a relatively well-conditioned regime with short data, used for comparison with the recent closed-loop DDPC controllers. The second is a weak-excitation regime, which serves as the main test since predictor uncertainty and regularization should matter most there.

We consider the sampled second-order system in \cite{innovation} and \cite{Breschi2023}, with the MPC setup matched to \cite{innovation}. The system matrices are
\begin{equation}
    \begin{aligned}
        A& = \begin{pmatrix}
            0.7326 & -0.0861     \\
            0.1722 & 0.9909
        \end{pmatrix}, \quad B  = \begin{pmatrix}
        0.0609 \\ 0.0064 \end{pmatrix} , \\
        C & = \begin{pmatrix}
            0 & 1.4142 \end{pmatrix}, \quad D=0 .
    \end{aligned}
\end{equation}
The process noise and measurement noise variance is $\Sigma_w = \sigma_w^2 \mathbb{I}_2$ and $\Sigma_v = \sigma_v^2$, respectively. The control objective is to track a sinusoidal reference signal $r(t) = \sin(2\pi t/75)$ under input and output constraints $-2 \leq u \leq 2, -2 \leq y \leq 2$. The past and future horizons for MPC are $L_p=10$ and $L_f=15$, respectively, and closed-loop performance is evaluated through the quadratic cost $J = \sum_{t=1}^{N_{\text{test}}} ||r(t)-y(t)||_Q^2 + ||u(t)||_R^2$ with $Q = 1$ and $R = 0.01$. The ARX-based controllers use order $n_a=n_b=10$, which is chosen slightly larger than the AIC given order when the sample size is large. Each test trajectory has length $N_{\text{test}} = 150$, and all reported results are based on $N_{MC} = 500$ Monte Carlo runs. 

\begin{figure}[t]
      \centering
      \includegraphics[width=0.48\textwidth]{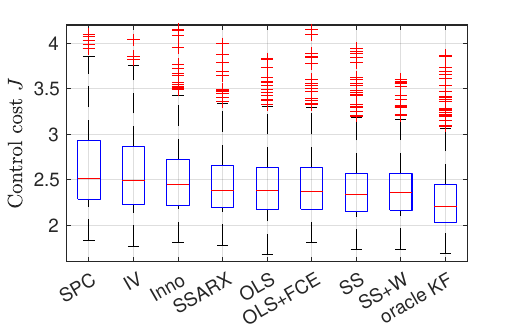}
      \caption{Closed-loop cost $J$ over 500 Monte Carlo runs in the informative-data regime. Since the identification problem is already reasonably well-conditioned, all ARX-based variants are competitive and only limited gains are obtained from additional kernel regularization.}
      \label{fig:1}
      \vspace{-13pt}
\end{figure}

\begin{figure*}[b]
\vspace{-13pt}
    \centering
    \includegraphics[width=\textwidth]{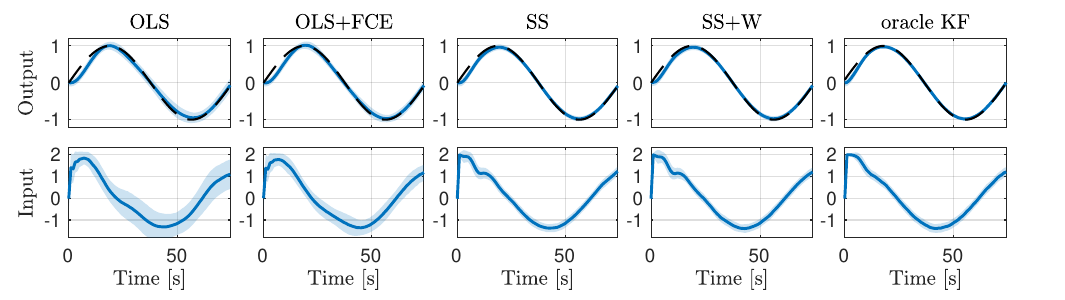}
    \caption{Mean and standard deviation of closed-loop trajectories in the weak-excitation regime, for the output (top row) and the input (bottom row). The figure illustrates not only nominal tracking but also variability reduction due to regularized identification.}
    \label{fig:traj}
\end{figure*}

As recent DDPC baselines, we use four external methods: SPC in \cite{SPC}, IV in \cite{IVDDPC}, Inno in \cite{innovation}, and SSARX in \cite{SSARX}. We compare them with several ARX-based variants derived from the proposed predictor framework and an oracle model-based benchmark (oracle KF). Here OLS denotes the plain ARX identification in (\ref{eq:arx_regress}), FCE adds the uncertainty penalty in the MPC cost as in (\ref{eq:FCE}), SS uses baseline SS-kernel regularization with EB tuning in (\ref{eq:theta_ss}), and SS+W adds the proposed sensitivity kernel in (\ref{eq:co_estimator}). We use the \textsc{Matlab} command \texttt{arxRegul} for SS kernel and EB tuning, and set $\mu=1$ for the un-normalized sensitivity kernel $\bar{W}$.

\begin{figure}[t]
      \centering
      \includegraphics[width=0.48\textwidth]{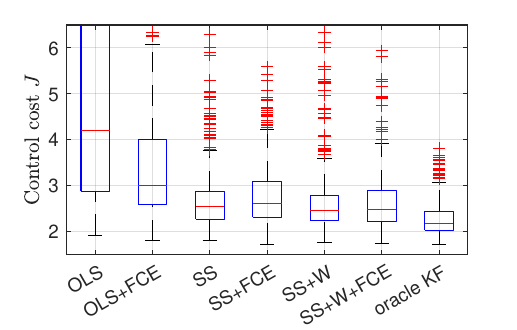}
      \caption{Closed-loop cost $J$ in the weak-excitation regime. Baseline SS regularization provides the main stabilization relative to OLS, while the proposed sensitivity shaping yields an additional improvement over SS.}
      \label{fig:cost}
      \vspace{-5pt}
\end{figure}


In the informative-data regime, the training trajectory has length $N_{\text{train}} = 150$, and is generated by a simple feedback law $u(t) = r_{\text{train}}(t)-y(t)$ driven by a square wave reference of amplitude $1$ and period $50$. 

Fig. \ref{fig:1} shows the closed-loop control performance. In this regime, all ARX-based variants are competitive with the DDPC baselines, and the additional gains from further regularization or FCE term are modest. This is consistent with the view that, once the identification problem is well conditioned, there is little harmful uncertainty left for the proposed sensitivity mechanism to exploit.



The weak-excitation regime is the main setting of interest, since it exposes the finite-data variance effects that motivate both regularized identification and sensitivity-guided shaping. The training data are again generated in closed-loop, this time by a sinusoidal reference of amplitude $1$ and period $75$, which yields less informative data and larger predictor uncertainty. In this case, the DDPC baselines frequently fail to yield valid closed-loop runs, so the detailed comparison in Fig. \ref{fig:traj} and Fig. \ref{fig:cost} is restricted to the ARX-based variants.

Under limited excitation, the role of regularization becomes much clearer. Fig. \ref{fig:traj} shows the mean and standard deviation of output and input trajectories for a representative subset of controllers, and Fig. \ref{fig:cost} reports the Monte Carlo cost distributions for the full ARX-based controller variants. 

In this regime, the dominant improvement comes from baseline SS regularization relative to OLS. The proposed sensitivity shaping then provides an additional improvement over SS, suggesting that the local Jacobian captures useful task-dependent structure beyond the identification-oriented prior alone. 

Adding the FCE-type term improves upon OLS but not further on already regularized predictors. A possible explanation is that the neglected cross term in (\ref{eq:FCE}) needs to remain small, which may be less accurate when the estimator is biased. Also, FCE might show an improvement in the tracking performance or worst-case control performance. 

To connect the closed-loop results in Fig. \ref{fig:cost} to the proposed mechanism, we also show the trace of the parameter posterior covariance $\operatorname{tr}(\Sigma_\theta)$ after identification in Table \ref{tab:1}. This reduction in the parameter posterior uncertainty $\Sigma_\theta$ is consistent with the observed performance improvement.

\begin{figure}[t]
    \centering
    \includegraphics[width=0.75\columnwidth]{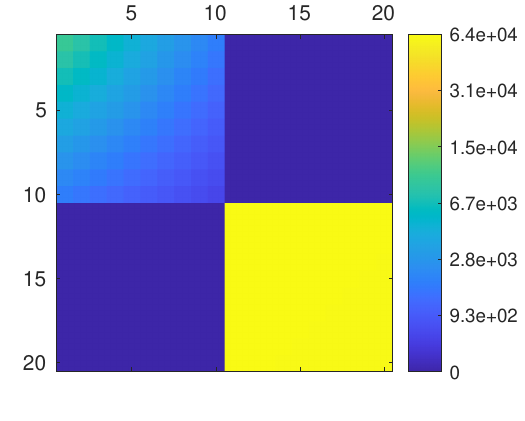}\\[-7mm]
    {\footnotesize (a) $K$}\\[1mm]
    \includegraphics[width=0.75\columnwidth]{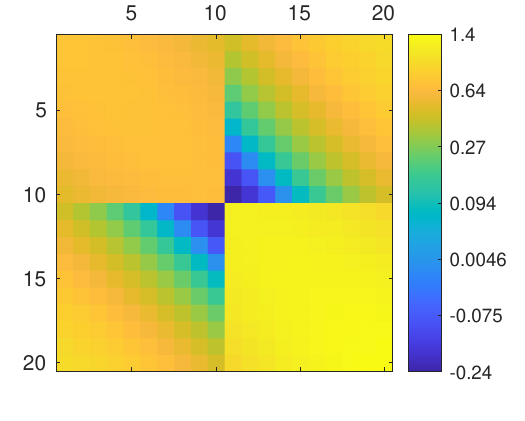}\\[-7mm]
    {\footnotesize (b) $\bar{W}_{\text{norm}}$}
    \caption{Heatmaps of (a) the empirical-Bayes tuned baseline SS kernel $K$ in (\ref{eq:K}), and (b) the normalized task-specific sensitivity kernel $\bar{W}_{\text{norm}}$ in (\ref{eq:normW}). Larger entries in $\bar{W}_{\text{norm}}$ indicate parameter directions with greater local influence on the weighted multi-step prediction error, and hence stronger candidates for shrinkage in the shaped estimator.}
    \label{fig:heatmap}
    \vspace{-15pt}
\end{figure}

Fig. \ref{fig:heatmap} illustrates how the baseline kernel and the normalized sensitivity matrix emphasize different parameter directions. The EB-tuned baseline SS kernel $K$ reflects identification-oriented prior structure, whereas the normalized sensitivity matrix $\bar{W}_{\text{norm}}$ indicates parameter directions that are locally more harmful for the weighted multi-step prediction error. Large values in $\bar{W}_{\text{norm}}$ indicate parameter directions that are locally more influential for the weighted multi-step prediction error, and therefore candidates for stronger shrinkage in the shaped kernel.




\begin{table}[t]
    \caption{Average posterior covariance $\operatorname{tr}(\Sigma_\theta)$ after identification}
    \vspace{-5pt}
    \label{tab:1}
    \centering
    \small
    \begin{tabular}{c|c|c}
        \hline
         OLS & SS & SS+W \\
        \hline
         0.0470$\pm$ 0.0028     & 0.0113 $\pm$ 0.0044     & 0.0019 $\pm$ 0.0012     \\
        \hline
    \end{tabular}
    \vspace{-13pt}
\end{table}

\section{Conclusions and Future Work} \label{sec:conclusion}
This paper developed a local first-order sensitivity analysis for the implicit structured ARX multi-step predictor. The main role of this analysis is to bridge linear predictor identification and control-relevant multi-step sensitivity. In the reported experiments, its clearest practical value lies in sensitivity-guided regularization under weak excitation: baseline SS regularization is the dominant source of improvement, while the proposed shaping provides an additional gain when predictor uncertainty remains significant. By contrast, direct FCE-style cost augmentation is less consistently beneficial in the present study. Future possibilities include investigating the tuning strategy for hyperparameters and control-oriented kernel design.




\section{Acknowledgment}


The authors acknowledge the use of generative AI (OpenAI ChatGPT) for language refinement and phrasing suggestions in parts of the manuscript. All technical content, interpretations, and final text were reviewed and approved by the authors.



   




\bibliographystyle{IEEEtran}
\bibliography{refs}






\end{document}